\documentclass[12pt]{article}
\usepackage{fullpage}
\usepackage[english]{babel}
\usepackage[utf8x]{inputenc}
\usepackage{latexsym}
\usepackage[titletoc,toc,title,page]{appendix}
\usepackage{cite}
\usepackage{amsfonts}
\usepackage{amsmath}
\usepackage{amsthm}
\usepackage{amssymb}
\usepackage{graphicx}
\usepackage{mathtools}
\usepackage{enumerate}
\usepackage{mathrsfs}
\usepackage{verbatim}
\usepackage[toc]{appendix}
\usepackage{color}
\numberwithin{equation}{section}

\DeclareMathOperator{\re}{Re}
\DeclareMathOperator{\im}{Im}
\newcommand\Backlund {B\"{a}cklund }

\usepackage{color}
\usepackage{cite}
\usepackage[width=\textwidth,font=small,labelfont=bf,
labelsep=endash]{caption}
\usepackage{titlesec}
\titleformat{\section}
  {\normalfont\fontsize{13}{16}\bfseries}{\thesection}{1em}{}
\titleformat{\subsection}
  {\normalfont\fontsize{11}{14}\bfseries}{\thesubsection}{1em}{}
\normalsize
\begin{document}
\title{\textbf{The Dressing Method as Non Linear Superposition in Sigma Models}}
\author{Dimitrios Katsinis$^{1,2}$, Ioannis Mitsoulas$^3$ and Georgios Pastras$^2$}
\date{\small $^1$Department of Physics, National and Kapodistrian University of Athens,\\University Campus, Zografou, Athens 15784, Greece\\
$^2$NCSR ``Demokritos'', Institute of Nuclear and Particle Physics,\\Aghia Paraskevi 15310, Attiki, Greece\\
$^3$Physics Division, National Technical University of Athens, Zografou Campus, Athens 15780, Greece\linebreak \vspace{8pt}
\texttt{dkatsinis@phys.uoa.gr, mitsoula@central.ntua.gr, pastras@inp.demokritos.gr}}

\vskip .5cm

\maketitle

\begin{abstract}
We apply the dressing method on the Non Linear Sigma Model (NLSM), which describes the propagation of strings on $\mathbb{R}\times \mathrm{S}^2$, for an arbitrary seed. We obtain a formal solution of the corresponding auxiliary system, which is expressed in terms of the solutions of the NLSM that have the same Pohlmeyer counterpart as the seed. Accordingly, we show that the dressing method can be applied without solving any differential equations. In this context a superposition principle emerges: The dressed solution is expressed as a non-linear superposition of the seed with solutions of the NLSM with the same Pohlmeyer counterpart as the seed.
\end{abstract}
\newpage
\tableofcontents
\newpage

\section{Introduction}
Integrability provides powerful tools for some special non - linear differential equations. An indicative example, that will occupy our interest, is provided by the equations of motion NLSMs on symmetric spaces. The equations of motion of the NLSM can be obtained as the compatibility condition of an auxiliary system, which is a linear system of first order differential equations. The dressing method \cite{Zakharov:1973pp,Zakharov:1980ty}  takes advantage of gauge transformations of the auxiliary system, which can be performed in a trivial manner, in order to construct new non-trivial solutions of the NLSM. The implementation of the dressing method relies on the existence of a known solution, which is called the seed solution. In order to obtain a dressed solution of the NLSM one has to solve the auxiliary system and impose appropriate constraints. The latter ensure that the dressed solution belongs to the symmetric space of the NLSM. Once this task is performed, a whole tower of solutions of the NLSM can be constructed algebraically.

Another aspect of integrability is related to the fact that the embedding equations of the solution of the NLSM in the target space, which is in turn embedded in an enhanced flat space, are in fact multicomponent generalizations of the sine-Gordon equation, the so called Symmetric Space Sine Gordon models (SSSGs). This is known as Pohlmeyer reduction\cite{Pohlmeyer:1975nb,Lund:1976ze,Miramontes:2008wt}. An important implication of the Pohlmeyer reduction is the fact that the equations of motion of the NLSM become linear, given a solution of the Pohlmeyer reduced theory. Notice that these linear equations do not provide the general solution of the NLSM, but only the solutions that correspond to the given Pohlmeyer counterpart. It is significant that the Pohlmeyer reduction is a many-to-one mapping. There is whole family of solutions of the NLSM that corresponds to the same Pohlmeyer field. In what follows, the term ``family" always refers to this set of solutions.

The SSSGs possess \Backlund transformations\cite{Bakas:1995bm}, which is the analogous to the existence of the dressing transformations for the NLSMs. These are sets of first order differential equations that interrelate different pairs of solutions of the SSSGs. It has been shown that dressing transformations are equivalent to \Backlund transformations on the Pohlmeyer reduced theory \cite{Hollowood:2009tw}. Solutions obtained by \Backlund transformations of the same seed solution can be combined algebraically using addition formulas in order to create new ones, see e.g.\cite{Park:1995np}. In order to construct new solutions algebraically there are first order differential equations to be solved. Regarding the NLSM, these are the auxiliary system, whereas on the Pohlmeyer reduced theory these are the \Backlund transformations. 

In \cite{Spradlin:2006wk} the dressing method was applied on the simplest possible seed, the BMN particle \cite{Berenstein:2002jq}, in order to produce the Giant Magnons \cite{Hofman:2006xt}. In our recent work \cite{Katsinis:2018ewd} we applied the dressing method on elliptic string solutions in $\mathbb{R}\times \mathrm{S}^2$, discussed in  \cite{Katsinis:2018zxi},  using a mapping between $\mathrm{S}^2$ and the coset $\mathrm{SO}(3)/\mathrm{SO}(2)$. The auxiliary system in this case was solved in a systematic way. This was achieved by incorporating an appropriate parametrization, which implements the dressing transformation through its effect on the identity element of the coset, along with special properties of the elliptic seeds. 

Motivated by the systematic solution of the auxiliary system for the elliptic seeds, in this work, we obtain a formal solution of the auxiliary system for an arbitrary seed. This formal solution is expressed in terms of a specific element of the family of the seed. This implies that the particular NLSM has a more fundamental property, which is a non-linear superposition rule. The dressing method is exactly the implementation of this non-linear superposition rule.

The structure of the paper is as follows: In section \ref{sec:strings_s2} we review basic elements of the NLSM that describes strings propagating on $\mathbb{R}\times \mathrm{S}^2$. In section \ref{sec:dressing} we discuss the dressing method and solve the auxiliary system for an arbitrary seed. In section \ref{sec:discussion} we discuss our results. There are also some appendices. Appendices \ref{sec:Normalization} and \ref{sec:remaining} contain some technical details on the derivation of the solution of the auxiliary system, while in appendix \ref{sec:dressing_factor} the construction of the simplest dressing factor is presented and the equivalence of the corresponding dressing transformation to the Pohlmeyer reduced theory is discussed. 

\section{Strings in $\mathbb{R}\times \mathrm{S}^2$}
\label{sec:strings_s2}
In this section, we introduce the conventions that are used throughout the text and discuss the equations that describe the propagation of strings on $\mathbb{R}\times \mathrm{S}^2$. Our convention for the world-sheet metric is $\mathrm{diag}(-1,1)$, while the light-cone world-sheet coordinates are defined as $\xi^\pm=\left(\xi^1\pm\xi^0\right)/2$. The last relation implies $\partial_\pm=\partial_1\pm\partial_0$. Additionally, we set the radius of $\textrm{S}^2$ equal to one. We consider $\textrm{S}^2$ as being embedded in a flat three dimensional space and denote the corresponding embedding functions as $\vec{X}$. The NLSM action reads
\begin{equation}\label{eq:action}
S=T\int d\xi^+d\xi^-\left[-\partial_+ X^0\partial_- X^0+\partial_+\vec{X}\cdot\partial_-\vec{X}+\nu\left(\vec{X}\cdot\vec{X}-1\right)\right],
\end{equation}
where $T$ is the string tension and $\nu$ is the Lagrange multiplier, which ensures that the string propagates on the sphere surface. The equations of motion read
\begin{align}
\partial_+\partial_-X^0&=0,\\
\partial_+\partial_-\vec{X}&=\nu\vec{X}.
\end{align}
The equation for the temporal component $X^0$ is trivially solved by
\begin{equation}
X^0(\xi^+,\xi^-)=f_+\left(\xi^+\right)+f_-\left(\xi^-\right).
\end{equation}
Taking advantage of the invariance under diffeomorphisms, we may select $\xi^\pm$ so that
\begin{equation}
f_\pm\left(\xi^\pm\right)=m_\pm\xi^\pm.
\end{equation}
This choice corresponds to a ``linear" gauge for the temporal component $X^0$, which is a generalization of the static gauge. By differentiating the geometric constraint $\vert \vec{X}\vert^2=1$ and using the equations of motion, we can determine the Lagrange multiplier, which reads
\begin{equation}
\nu=-\left(\partial_+\vec{X}\right)\cdot\left(\partial_-\vec{X}\right).
\end{equation}

The action \eqref{eq:action} is supplemented by the Virasoro constraints, which read
\begin{equation}\label{eq:def_Vir}
\partial_\pm\vec{X}\cdot\partial_\pm\vec{X}=m^2_\pm.
\end{equation}
The embedding of the string in the enhanced flat space of $\textrm{S}^2$, which is $\mathbb{R}^3$, is controlled by a single degree of freedom, the Pohlmeyer field $a$, which is defined so that
\begin{equation}\label{eq:def_Pohl}
\partial_+\vec{X}\cdot\partial_-\vec{X}=m_+ m_-\cos a.
\end{equation}
The standard Pohlmeyer reduction implies that $ a$ satisfies the sine-Gordon equation \cite{Pohlmeyer:1975nb}. In our conventions the latter reads
\begin{equation}\label{eq:SG}
\partial_+\partial_- a=-m_+ m_-\sin a,
\end{equation}
where $m_+m_-<0,$ so that $\xi^0$ is the time-like world-sheet coordinate and $\xi^1$ the space-like one. For details on the Pohlmeyer reduction without gauge fixing $f_+$ and $f_-$ see \cite{Katsinis:2018zxi}. Taking advantage of the Pohlmeyer field, the equations of motion of the NLSM read
\begin{equation}\label{eq:NLSM_EOM}
\partial_+\partial_-\vec{X}=-m_+m_-\cos a\vec{X}.
\end{equation}
For a given a solution of the Pohlmeyer reduced theory, i.e. the sine-Gordon equation, these equations are linear.

We introduce the usual spherical coordinates: the polar angle $\theta$ and the azimuthal angle $\phi$, where $\theta=0$ corresponds to the z-axis. Then, the equations of motion read
\begin{align}
\partial_0\left[\sin^2\theta\partial_0\phi\right]&=\partial_1\left[\sin^2\theta\partial_1\phi\right],\\
\partial_0^2\theta-\cos\theta\sin\theta\left(\partial_0\phi\right)^2&=\partial_1^2\theta-\cos\theta\sin\theta\left(\partial_1\phi\right)^2,
\end{align}
while the Virasoro constraints assume the form
\begin{equation}
\left(\partial_1\theta \pm \partial_0\theta\right)^2+\sin^2\theta\left(\partial_1\phi \pm \partial_0\phi\right)^2=m^2_\pm.
\end{equation}
Finally, the coordinates of the string are related to the Pohlmeyer field by
\begin{equation}
\left(\partial_1\theta\right)^2 - \left(\partial_0\theta\right)^2+\sin^2\theta\left[\left(\partial_1\phi\right)^2 - \left(\partial_0\phi\right)^2\right]=m_+m_-\cos a.
\end{equation}
This relation does not determine the sign of the Pohlmeyer field. We fix it by demanding 
\begin{equation}\label{eq:rule_sin_Pohl}
\vec{X}\cdot\left(\partial_+\vec{X}\times\partial_-\vec{X}\right)=2\sin\theta\left[\partial_0\theta\partial_1\phi-\partial_1\theta\partial_0\phi\right]=m_+m_-\sin a.
\end{equation}
Combining the above, we obtain the following expressions that will be used extensively in what follows:
\begin{align}
\left(\partial_0\theta\right)^2+\sin^2\theta\left(\partial_0\phi\right)^2&=\frac{m_+^2}{4}+\frac{m_-^2}{4}-\frac{m_+m_-}{2}\cos a,\label{eq:rule00}\\
\left(\partial_1\theta\right)^2+\sin^2\theta\left(\partial_1\phi\right)^2&=\frac{m_+^2}{4}+\frac{m_-^2}{4}+\frac{m_+m_-}{2}\cos a,\label{eq:rule11}\\
\partial_0\theta\partial_1\theta+\sin^2\theta\partial_0\phi\partial_1\phi&=\frac{m_+^2}{4}-\frac{m_-^2}{4}.\label{eq:rule01}
\end{align}
It is important to point out that the Pohlmeyer reduced theory depends only on the product $m_+m_-$. The Pohlmeyer reduction is a many-to-one mapping. For each solution of the Pohlmeyer reduced theory, there is a whole family of NLSM solutions\footnote{This family is an associate (Bonnet) family of world-sheets.}, which corresponds to the same product $m_+m_-$. Its members are parametrized by the ratio $m_+/m_-$, see e.g. \cite{Katsinis:2018zxi}.

\section{Dressed Strings in  $\mathbb{R}\times S^2$}
\label{sec:dressing}
The dressing method enables us to construct new solutions of a NLSM once we are given a solution of the latter. We refer to this solution as seed solution. Given the seed solution we may obtain a new solution of the NLSM by solving a pair of first order equations, which are called the auxiliary system. This is considered a simpler task than solving the original equations of motion, which are non-linear and second order. We will not only verify this statement, but we will obtain the solution of the auxiliary system, which corresponds to strings in $\mathbb{R}\times S^2$, for an arbitrary seed.

The auxiliary system reads
\begin{equation}\label{eq:auxiliary_system}
\partial_\pm\Psi(\lambda)=\frac{1}{1\pm\lambda}\left(\partial_\pm g\right)g^{-1}\Psi(\lambda),
\end{equation}
where $\lambda$ is the spectral parameter, which is in general complex. The seed solution $X$ is mapped to an element of the coset $SO(3)/SO(2)$, which is denoted as $g$. The compatibility relation $\partial_+\partial_-\Psi=\partial_-\partial_+\Psi,$ which ensures the local existence of a solution of the auxiliary system, implies that $g$ obeys the equations of motion $\partial_+\left(\left(\partial_-g\right)g^{-1}\right)+\partial_-\left(\left(\partial_+g\right)g^{-1}\right)=0$. The normalization of $\Psi(\lambda)$ is fixed as
\begin{equation}\label{eq:Psi_initial_condition}
\Psi(0)=g.
\end{equation}
The main idea of the dressing method is the fact that a gauge transformation of the auxiliary field
\begin{equation}
\Psi^\prime(\lambda)=\chi(\lambda)\Psi(\lambda),
\end{equation} 
corresponds to a new, non-trivial element of the coset, namely
\begin{equation}
g^\prime=\chi(0)g,
\end{equation}
which is associated to a new string solution, via the inverse mapping. We refer the reader to \cite{Hollowood:2009tw,Katsinis:2018ewd} for more details on the dressing method.

The mapping from $\mathbb{R}^3,$ the enhanced space of $\mathrm{S}^2$, to the coset $\mathrm{SO}(3)/\mathrm{SO}(2),$ that is used, is
\begin{equation}\label{eq:g_mapping}
g=J\left(I-2XX^T\right),\qquad J=\left(I-2X_0X_0^T\right),
\end{equation}
where $X_0$ is a constant vector and $X^TX=X_0^TX_0=1.$ It is easy to show that $\left(I-2XX^T\right)^2=I,$ for any unit norm vector $X$, which implies that
\begin{equation}\label{eq:g_constraints_1}
gJ gJ=I,\qquad
g^T=g^{-1}.
\end{equation}
In addition, $g$ is real, i.e.
\begin{equation}
\bar{g}=g.\label{eq:g_constraints_2}
\end{equation}
Thus, $g$ is indeed an element of the coset $\mathrm{SO}(3)/\mathrm{SO}(2)$. On a more formal basis, starting with the group $\mathrm{SL}(2;\mathbb{C})$, the coset can by constructed using the following involutions
\begin{align}
\sigma_1\left(g\right)&=(g^\dagger)^{-1},\label{eq:inv_1}\\
\sigma_2\left(g\right)&=JgJ,\label{eq:inv_2}\\
\sigma_3\left(g\right)&=\bar{g}.\label{eq:inv_3}
\end{align}
Demanding invariance under the first involution restricts $g$ to $\mathrm{SU}(3)$. Setting $\sigma_2\left(g\right)=g^{-1}$ restricts $g$ further, to the coset $\mathrm{SU}(3)/\mathrm{U}(2)$. Finally, invariance under the last involution implies that $g$ is an element of the coset $\mathrm{SO}(3)/\mathrm{SO}(2)$. Applying the same involutions on the auxiliary system \eqref{eq:auxiliary_system} implies that the transformed $\Psi(\xi^0,\xi^1;\lambda)$ must belong to the set of solutions of the auxiliary system. The latter is generated by the right multiplication with a constant matrix of a given solution; in our discussion this solution is $\Psi(\xi^0,\xi^1;\lambda)$. Thus, the following constraints must be imposed\footnote{Equation \eqref{eq:Psi_inv} corresponds to the action of both involutions \eqref{eq:inv_1} and \eqref{eq:inv_3}.}:
\begin{align}
\Psi(\lambda)m_1(\lambda)&=\left(\Psi(\lambda)^T\right)^{-1},\label{eq:Psi_inv}\\
\Psi(\lambda)m_2(\lambda)&=gJ\Psi(1/\lambda)J,\label{eq:Psi_cos}\\
\Psi(\lambda)m_3(\lambda)&=\overline{\Psi(\bar{\lambda})}.\label{eq:Psi_real}
\end{align}
The matrices $m_i$ themselves are subject to constraints, which stem from the fact that the involutions satisfy $\sigma^2=I$. In particular, they obey
\begin{align}
m_1(\lambda)=m_1^T(\lambda),\label{eq:constraint_m_inv}\\
m_2(\lambda)J m_2(1/\lambda) J=I,\label{eq:constraint_m_cos}\\
m_3(\lambda)\bar{m}_3(\bar{\lambda})=I.\label{eq:constraint_m_real}
\end{align}
In addition, since $\Psi(0)=g$ the matrices $m_1$ and $m_3$ must reduce to the identity matrix for $\lambda=0$, i.e.
\begin{equation}
m_1(0)=m_3(0)=I.\label{eq:constraint_m_0}
\end{equation}
These matrices are related to the so called reduction group \cite{Zakharov:1973pp,Mikhailov:1981us}. As we will show subsequently, the dressed string solution is not affected by the choice of these matrices. 

\subsection{The  Auxiliary System}
Having set the framework, we are ready to implement the dressing method. We follow the approach introduced in \cite{Katsinis:2018ewd}, parametrizing the seed as a rotation of a constant reference vector. In order to proceed we change the coordinates of the auxiliary system \eqref{eq:auxiliary_system} from the left- and right-moving coordinates $\xi^{\pm}$ to $\xi^0$ and $\xi^1$. The auxiliary system assumes the form
\begin{equation}
\partial_i\Psi(\lambda)=\left[\left(\tilde{\partial}_i g\right)g^{-1}\right]\Psi(\lambda),
\end{equation}
where $i=0,1$ and
\begin{equation}
\tilde{\partial}_{0/1}=\frac{1}{1-\lambda^2}\partial_{0/1}-\frac{\lambda}{1-\lambda^2}\partial_{1/0}.
\end{equation}
We express the seed solution of the NLSM as
\begin{equation}
X= U \hat{X},
\end{equation}
where $\hat{X}$ is a constant vector and $U$ is a rotation matrix. The element of the coset that corresponds to the seed solution can be expressed as
\begin{equation}\label{eq:ghat_definition}
g=J U J \hat{g}U^T,
\end{equation}
where 
\begin{equation}\label{eq:ghat_mapping}
\hat{g}:=J\left(I-2\hat{X}\hat{X}^T\right).
\end{equation}
Notice that $\hat{g}$ is an element of the coset. In a similar manner, we define $\hat{\Psi}$ as
\begin{equation}\label{eq:psihat_definition}
\Psi=J U J\hat{\Psi}.
\end{equation}
The auxiliary system assumes the form
\begin{equation}
\partial_i\hat{\Psi}=\left\{J U^T \left[\left(\tilde{\partial}_i-\partial_i\right)U\right]J-\hat{g}U^T\left[\tilde{\partial}_iU\right]\hat{g}^T+\left[\tilde{\partial}_i\hat{g}\right]\hat{g}^T\right\}\hat{\Psi}.
\end{equation}
We select $X_0$ to be the unit norm vector along the z axis, i.e.
\begin{equation}
X_0=\begin{pmatrix}
0 \\ 0 \\ 1
\end{pmatrix},
\end{equation}
so that $J=\mathrm{diag}(1,1,-1)$.  Moreover, the matrix $U$ can be selected so that $\hat{X}=X_0$. Thus, $\hat{g}$ becomes the identity element of the coset and the equations of the auxiliary system assume the form
\begin{equation}
\partial_i\hat{\Psi}=\left\{J U^T \left[\left(\tilde{\partial}_i-\partial_i\right)U\right]J-U^T\left[\tilde{\partial}_iU\right]\right\}\hat{\Psi},
\end{equation}
while the normalization \eqref{eq:Psi_initial_condition} reduces to
\begin{equation}\label{eq:Psi_hat_initial_condition}
\hat{\Psi}(0)=U^T.
\end{equation}
In addition, the constraints \eqref{eq:Psi_inv}, \eqref{eq:Psi_cos} and \eqref{eq:Psi_real} for $\Psi$, imply that $\hat{\Psi}$ is subject to the following constraints:
\begin{align}
\hat{\Psi}(\lambda)m_1(\lambda)&=\left(\hat{\Psi}(\lambda)^T\right)^{-1},\label{eq:Psi_hat_inv}\\
\hat{\Psi}(\lambda)m_2(\lambda)&=J\hat{\Psi}(1/\lambda)J,\label{eq:Psi_hat_cos}\\
\hat{\Psi}(\lambda)m_3(\lambda)&=\bar{\hat{\Psi}}(\bar{\lambda}).\label{eq:Psi_hat_real}
\end{align}
We express the matrix $U$ as $U=U_2U_1$, where
\begin{equation}
U_1=\begin{pmatrix}
\cos\theta & 0 & \sin\theta \\
0 & 1 & 0 \\
- \sin\theta & 0 & \cos\theta
\end{pmatrix}, \qquad U_2= \begin{pmatrix}
\cos\phi & -\sin\phi & 0\\
\sin\phi & \cos\phi & 0\\
0 & 0 & 1
\end{pmatrix}.\label{eq:def_U}
\end{equation}
The auxiliary system can by expressed as
\begin{equation}\label{eq:auxiliary_psi_hat}
\partial_i\hat{\Psi}=\left(t_i^jT_j\right)\hat{\Psi},
\end{equation}
where $T_j$ are the generators of the group $\mathrm{SO}(3)$:
\begin{equation}
T_1=\begin{pmatrix}
0 & 0 & 0\\
0 & 0 & -1\\
0 & 1 & 0
\end{pmatrix},\quad
T_2=\begin{pmatrix}
 0 & 0 & 1\\
 0 & 0 & 0\\
-1 & 0 & 0
\end{pmatrix},\quad
T_3=\begin{pmatrix}
0 & -1 & 0\\
1 & 0 & 0\\
0 & 0 & 0
\end{pmatrix}.\label{eq:generators_T}
\end{equation}
It is a matter of algebra to show that
\begin{align}
t_{0/1}^1&=\sin\theta\left(\frac{1+\lambda^2}{1-\lambda^2}\partial_{0/1}\phi- \frac{2\lambda}{1-\lambda^2}\partial_{1/0}\phi\right),\\
t_{0/1}^2&=-\frac{1+\lambda^2}{1-\lambda^2}\partial_{0/1}\theta+\frac{2\lambda}{1-\lambda^2}\partial_{1/0}\theta,\\
t_{0/1}^3&=-\cos\theta\partial_{0/1}\phi.
\end{align}
For later convenience we define
\begin{equation}
\vec{t}_i:=\begin{pmatrix}
t_i^1\\
t_i^2\\
t_i^3
\end{pmatrix},
\qquad 
\vec{\tau}_i:=\vec{t}_i-\left(\vec{X}_0\cdot\vec{t}_i\right)\vec{X}_0=\begin{pmatrix}
t_i^1\\
t_i^2\\
0
\end{pmatrix}.\label{eq:def_vectors}
\end{equation}
Notice that
\begin{equation}
\frac{d}{d\lambda}\vec{t}_0=-\frac{2\lambda}{1-\lambda^2}\vec{\tau}_1,\qquad\frac{d}{d\lambda}\vec{t}_1=-\frac{2\lambda}{1-\lambda^2}\vec{\tau}_0\label{eq:der_lambda}.
\end{equation}
Under the inversion of $\lambda\rightarrow 1/\lambda$ the quantities $\vec{\tau}_i$ and $t_i^3$ have the following parity properties
\begin{align}\label{eq:tau_inv}
\vec{\tau}_i(1/\lambda)&=-\vec{\tau}_i(\lambda),\qquad t_i^3(1/\lambda)=t_i^3(\lambda).
\end{align}
In addition, all quantities are real functions of the complex spectral parameter, i.e.
\begin{align}\label{eq:t_reality}
\bar{\vec{t}}_i(\bar{\lambda})&=\vec{t}(\lambda).
\end{align}
The derivatives of $\vec{t}_i$ and $\vec{\tau}_i$ obey the following algebra
\begin{align}
\partial_1\vec{t}_0-\partial_0\vec{t}_1&=\vec{t}_1\times\vec{t}_0,\label{eq:der_t}\\
\partial_1\vec{\tau}_1-\partial_0\vec{\tau}_0&=\vec{\tau}_0\times\vec{t}_0+\vec{t}_1\times\vec{\tau}_1.\label{eq:der_tau}
\end{align}
Notice that \eqref{eq:der_tau} can be obtained from \eqref{eq:der_t} using \eqref{eq:der_lambda}.

Moreover, it is straightforward to show that:
\begin{align}
\vert\vec{\tau}_0\vert^2&=\frac{m_+^2}{4}\left(\frac{1-\lambda}{1+\lambda}\right)^2+\frac{m_-^2}{4}\left(\frac{1+\lambda}{1-\lambda}\right)^2-\frac{m_+m_-}{2}\cos a,\label{eq:t00}\\
\vert\vec{\tau}_1\vert^2&=\frac{m_+^2}{4}\left(\frac{1-\lambda}{1+\lambda}\right)^2+\frac{m_-^2}{4}\left(\frac{1+\lambda}{1-\lambda}\right)^2+\frac{m_+m_-}{2}\cos a,\label{eq:t11}\\
\vec{\tau}_0\cdot\vec{\tau}_1&=\frac{m_+^2}{4}\left(\frac{1-\lambda}{1+\lambda}\right)^2-\frac{m_-^2}{4}\left(\frac{1+\lambda}{1-\lambda}\right)^2.\label{eq:t01}
\end{align}
The careful reader will that recognize these relations are identical to \eqref{eq:rule00}, \eqref{eq:rule11} and \eqref{eq:rule01} upon the substitutions $\partial_i\vec{X}\rightarrow\vec{\tau}_i$ and
\begin{equation}\label{eq:m_lambda}
m^2_\pm\rightarrow m^2_\pm \left(\frac{1\mp\lambda}{1\pm\lambda}\right)^2.
\end{equation}
This fact will be crucial in what follows. In addition, one may obtain
\begin{equation}\label{eq:cross_01}
\vec{\tau}_0\times\vec{\tau}_1=\frac{1}{2}m_+m_-\sin a\vec{X}_0,
\end{equation}
which is analogous to \eqref{eq:rule_sin_Pohl}.

Finally, the expressions of $U_1$ and $U_2$, which are given by \eqref{eq:def_U}, imply that the condition \eqref{eq:Psi_hat_initial_condition} assumes the form
\begin{equation}\label{eq:Psi_hat_initial_condition_explicit}
\hat{\Psi}(0)=\begin{pmatrix}
\cos\theta\cos\phi & \cos\theta\sin\phi & -\sin\phi \\
-\sin\phi & \cos\phi & 0 \\
\sin\theta\cos\phi & \sin\theta\sin\phi & \cos\theta
\end{pmatrix}.
\end{equation}
\subsection{The Solution of the Auxiliary System}
The auxiliary system \eqref{eq:auxiliary_psi_hat} comprises of three independent, identical, pairs of equations, one for each column of  $\hat{\Psi}$, which we denote as\footnote{In this notation $\hat{\Psi}=\begin{pmatrix}\vec{\hat{\Psi}}_1 & \vec{\hat{\Psi}}_2 & \vec{\hat{\Psi}}_3\end{pmatrix}$.} $\vec{\hat{\Psi}}_j$. In particular, each column obeys the equations 
\begin{equation}\label{eq:auxialiary_vector_psi_hat}
\partial_i\vec{\hat{\Psi}}_j=\vec{t}_i\times\vec{\hat{\Psi}}_j,
\end{equation}
where $j=1,2,3.$ Let us consider the inner product of two arbitrary solutions of this system of equations. It is straightforward to show that
\begin{equation}\label{eq:d_inner_product}
\partial_i\left[\vec{\hat{\Psi}}_j\cdot\vec{\hat{\Psi}}_k\right]=\left(\vec{t}_i\times\vec{\hat{\Psi}}_j\right)\cdot\vec{\hat{\Psi}}_k+\vec{\hat{\Psi}}_j\cdot\left(\vec{t}_i\times\vec{\hat{\Psi}}_k\right)=0.
\end{equation}
This proves that the constraint \eqref{eq:Psi_hat_inv}, which implies $\vec{\hat{\Psi}}_j\cdot\vec{\hat{\Psi}}_k=\left(m^{-1}_1(\lambda)\right)_{jk}$, is compatible with the equations of the auxiliary system\footnote{We remind the reader that the matrix $m_1(\lambda)$, is symmetric due to \eqref{eq:constraint_m_inv}.}. The system \eqref{eq:auxialiary_vector_psi_hat} has three linearly independent solutions. For some given $\xi^0$ and $\xi^1$ we may specify linear combinations of these solutions, which we denote as $\vec{\hat{V}}_j$ that form an orthonormal basis.  Due to the linearity of the equations \eqref{eq:auxialiary_vector_psi_hat}, $\vec{\hat{V}}_j$ satisfy
\begin{equation}\label{eq:auxialiary_vector}
\partial_i\vec{\hat{V}}_j=\vec{t}_i\times\vec{\hat{V}}_j.
\end{equation}
Then, equation \eqref{eq:d_inner_product} implies that these vectors form an orthonormal basis for any $\xi^0$ and $\xi^1$, i.e.
\begin{equation}\label{eq:con_orthonormal}
\vec{\hat{V}}_j\cdot\vec{\hat{V}}_k=\delta_{jk}.
\end{equation}
We will solve \eqref{eq:auxialiary_vector} by projecting it on linear independent directions, namely $\vec{\hat{V}}_j$, $\vec{X}_0$ and $\vec{X}_0\times\vec{\tau}_i$. Obviously, the equations obtained by the projection of \eqref{eq:auxialiary_vector} along $\vec{\hat{V}}_j$, are redundant, since they are equivalent to the constraint \eqref{eq:con_orthonormal}.

Recognizing that the third components of $\vec{\hat{V}}_j$ are special will enable us to solve the rest of the equations of the auxiliary system, as well as the constraints. This is due to the fact that $\vec{X}_0$ is parallel to the third axis. These components obey the same equations of motion as the embedding functions of the string solution \eqref{eq:NLSM_EOM}, i.e
\begin{equation}\label{eq:V_3_eom}
\partial_1^2\hat{V}_j^3-\partial_0^2\hat{V}_j^3=-m_+m_-\cos a \hat{V}_j^3.
\end{equation}

Let us prove this statement. The auxiliary system \eqref{eq:auxialiary_vector} implies that
\begin{multline}
\partial_1^2 \hat{V}_j^3-\partial_0^2 \hat{V}_j^3= \\ \left[\vec{X}_0\times\left(\partial_1 \vec{\tau}_1-\partial_0 \vec{\tau}_0\right)\right]\cdot\vec{\hat{V}}_j+\left[\vec{t}_1\times\left[\vec{t}_1\times\vec{\hat{V}}_j\right]\right]\cdot \vec{X}_0-\left[\vec{t}_0\times\left[\vec{t}_0\times\vec{\hat{V}}_j\right]\right]\cdot \vec{X}_0.
\end{multline}
Taking the equation \eqref{eq:der_tau}, as well as \eqref{eq:def_vectors}, into account, it is easy to show that
\begin{equation}
\partial_1^2 \hat{V}_j^3-\partial_0^2 \hat{V}_j^3=-\left(\vert\vec{\tau}_1\vert^2-\vert\vec{\tau}_0\vert^2\right)\hat{V}_j^3.
\end{equation}
It is important that the component of $\vec{\hat{V}}_j$, which is parallel to $\vec{X}_0$, namely $\hat{V}_j^3$, obeys a second order equation that is \emph{decoupled}, i.e. it does not contain the other components. Moreover, equations \eqref{eq:t00} and \eqref{eq:t11} trivially imply that this relation assumes the form of \eqref{eq:V_3_eom}.

It is evident that we have to single out $\hat{V}_j^3$ and use the auxiliary system in order to express the other two components $\hat{V}_j^1$ and  $\hat{V}_j^2$ in terms of the former. The projection of the auxiliary system \eqref{eq:auxialiary_vector} on the direction of $\vec{X}_0$ reads
\begin{equation}
\partial_i\hat{V}_j^3=\left[\vec{X}_0\times\vec{\tau}_i\right]\cdot\vec{\hat{V}}_j,
\end{equation}
since $\vec{X}_0\times\vec{t}_i=\vec{X}_0\times\vec{\tau}_i$. This implies that the column $\vec{\hat{V}}_j$ has the following form
\begin{equation}\label{eq:column_form}
\vec{\hat{V}}_j= \frac{\vec{\tau}_1}{\left(\vec{\tau}_0\times\vec{\tau}_1\right)\cdot\vec{X}_0}\partial_0 \hat{V}_j^3- \frac{\vec{\tau}_0}{\left(\vec{\tau}_0\times\vec{\tau}_1\right)\cdot\vec{X}_0}\partial_1 \hat{V}_j^3+\hat{V}_j^3\vec{X}_0.
\end{equation}
Thus, the components $\hat{V}_j^3$ completely specify the solution $\vec{\hat{V}}_j$.

Finally, we may obtain another pair of independent equations by projecting the auxiliary system $\eqref{eq:auxialiary_vector}$ on $\vec{X}_0\times\vec{\tau}_i$. After some simple algebraic manipulations, this yields
\begin{equation}\label{eq:last_eq}
\partial^2_i \hat{V}_j^3=\left(\partial_i\vec{\tau}_i-\vec{t}_i\times\vec{\tau}_i\right)\cdot\left[\vec{\hat{V}}_j\times\vec{X}_0\right]-\vert\vec{\tau}_i\vert^2 \hat{V}_j^3.
\end{equation}
In virtue of \eqref{eq:der_tau}, the difference of this equation for $i=0$ and $i=1$ is trivially equation \eqref{eq:V_3_eom}. Thus, the vector $\vec{\hat{V}}_j$ has the form of equation \eqref{eq:column_form}, where $\hat{V}_j^3$ obeys equations \eqref{eq:V_3_eom} and \eqref{eq:last_eq} for either values of $i$. In addition one has to impose the constraints \eqref{eq:con_orthonormal}.

We continue the discussion inspired by the latter. As a direct consequence of equation \eqref{eq:con_orthonormal} it is true that $\sum_j\left(\hat{V}^3_j\right)^2=1$. Therefore, having at the back of our mind equation \eqref{eq:Psi_hat_initial_condition_explicit} we may define
\begin{align}
\hat{V}^3_1&=\sin\Theta\cos\Phi,\\
\hat{V}^3_2&=\sin\Theta\sin\Phi,\\
\hat{V}^3_3&=\cos\Theta,
\end{align}
where $\Theta=\Theta\left(\xi^0,\xi^1;\lambda\right)$ and $\Phi=\Phi\left(\xi^0,\xi^1;\lambda\right)$, will be specified by the various equations and constraints. It is obvious that the condition \eqref{eq:Psi_hat_initial_condition_explicit} is equivalent to the fact that for $\lambda=0$ these functions should reduce to the coordinates of the seed solution, i.e.
\begin{equation}\label{eq:initial_condition_theta_phi}
\Theta\vert_{\lambda=0}=\theta,\qquad \Phi\vert_{\lambda=0}=\phi.
\end{equation}
So far, we know that the functions $\Theta$ and $\Phi$ could decribe a solution of the NLSM, which has the same Pohlmeyer counterpart as the seed solution. We turn to the condition that $\vec{\hat{V}}_j$ should form an orthonormal basis as suggested by \eqref{eq:con_orthonormal}. This condition allows us to obtain the analogous of the Virasoro constraints, which are obeyed by $\Theta$ and $\Phi$. We derive them in appendix \ref{sec:Normalization}. They read
\begin{align}
\left(\partial_0\Theta\right)^2+\sin^2\Theta\left(\partial_0\Phi\right)^2&=\vert\vec{\tau}_0\vert^2=\frac{m_+^2}{4}\frac{\left(1-\lambda\right)^2}{\left(1+\lambda\right)^2}+\frac{m_-^2}{4}\frac{\left(1+\lambda\right)^2}{\left(1-\lambda\right)^2}-\frac{m_+m_-}{2}\cos a,\label{eq:Vir_00_lambda}\\
\left(\partial_1\Theta\right)^2+\sin^2\Theta\left(\partial_1\Phi\right)^2&=\vert\vec{\tau}_1\vert^2=\frac{m_+^2}{4}\frac{\left(1-\lambda\right)^2}{\left(1+\lambda\right)^2}+\frac{m_-^2}{4}\frac{\left(1+\lambda\right)^2}{\left(1-\lambda\right)^2}+\frac{m_+m_-}{2}\cos a,\label{eq:Vir_11_lambda}\\
\partial_0\Theta\partial_1\Theta+\sin^2\Theta\partial_0\Phi\partial_1\Phi&=\vec{\tau}_0\cdot\vec{\tau}_1=\frac{m_+^2}{4}\frac{\left(1-\lambda\right)^2}{\left(1+\lambda\right)^2}-\frac{m_-^2}{4}\frac{\left(1+\lambda\right)^2}{\left(1-\lambda\right)^2}.\label{eq:Vir_01_lambda}
\end{align}
In appendix \ref{sec:remaining}, we show that equation \eqref{eq:last_eq} is satisfied without any further constraints on $\Theta$ and $\Phi$.

Therefore, the triplet which is composed by the third components of the vectors $\vec{\hat{V}}_j$ obeys:
\begin{enumerate}
\item The normalization $\sum_j\left(\hat{V}^3_j\right)^2=1$, which is analogous to the geometric constraint $\vert\vec{X}\vert^2=1$ that defines the $S^2$. This justifies the definition of $\Theta$ and $\Phi$ in the same fashion as $\theta$ and $\phi$  in the original NLSM.\\
\item Equations \eqref{eq:Vir_00_lambda}, \eqref{eq:Vir_11_lambda} and \eqref{eq:Vir_01_lambda} which are identical to equations \eqref{eq:rule00}, \eqref{eq:rule11} and \eqref{eq:rule01}, upon the substitution \eqref{eq:m_lambda}. It is important that this transformation leaves the product $m_+m_-$ invariant. This implies that triplet $\hat{V}^3_j$ obeys the same ``Virasoro'' constraints as the seed but with different constants $m_\pm$ and it has the same ``Pohlmeyer counterpart'' as the seed solution.
\item The equation \eqref{eq:V_3_eom} which is identical to the equations of motion \eqref{eq:NLSM_EOM} obeyed by the components of the original seed solution with given Pohlmeyer counterpart.
\end{enumerate}

Thus, following the discussion at the end of section \ref{sec:strings_s2}, the triplet $\hat{V}^3_j$ is given by \emph{the member of the family of the seed which corresponds to the ratio }
\begin{equation}
\frac{m_+}{m_-}\left(\lambda\right)=\left(\frac{1+\lambda}{1-\lambda}\right)^2\frac{m_+}{m_-}.\label{eq:m_of_lambda}
\end{equation}
However, since $\lambda$ is in general complex, one is not restricted to the real solutions of the family of the seed, but rather to its analytic continuation.

Obviously, for $\lambda=0$, equations \eqref{eq:Vir_00_lambda}, \eqref{eq:Vir_11_lambda} and \eqref{eq:Vir_01_lambda} reduce to the relevant equations of the seed solution. One may be tempted to regard this as the fact that this  ``virtual'' solution of the NLSM reduces to the seed one, yet this is true up to global rotations. To ensure that no such global rotation is involved, so that the condition \eqref{eq:initial_condition_theta_phi} is satisfied, one has to employ \eqref{eq:m_lambda} directly to the coordinates of the seed solution.

Let $\hat{V}(\lambda)$ be the matrix, whose columns are the three orthonormal solutions $\vec{\hat{V}}_j$ of the system \eqref{eq:auxialiary_vector} that we constructed above. Taking into account the freedom of the right multiplication of a solution of \eqref{eq:auxiliary_psi_hat} with a constant matrix $C(\lambda)$, we consider the whole class of solutions of the auxiliary system
\begin{equation}\label{eq:definition_C}
\hat{\Psi}(\lambda)=\hat{V}(\lambda) C(\lambda).
\end{equation}
Obviously, equation \eqref{eq:con_orthonormal} implies
\begin{equation}\label{eq:V_Orth}
\hat{V}^T(\lambda)=\hat{V}^{-1}(\lambda).
\end{equation}
The equation \eqref{eq:tau_inv}, implies that the matrix $\hat{V}$ transforms under the inversion of $\lambda$ as
\begin{equation}\label{eq:V_Coset}
\hat{V}\left(1/\lambda\right)=-J \hat{V}\left(\lambda\right)M\left(\lambda\right),
\end{equation}
where the matrix $M$ represents the transformation of $\hat{V}_j^3$ under $\lambda\rightarrow 1/\lambda$. Since $\hat{V}_j^3(\lambda)$ satisfy the equations of motion \eqref{eq:V_3_eom}, it implies that $\hat{V}_j^3(1/\lambda)$ belongs to the set of solutions of this equation. In addition $\hat{V}_j^3(1/\lambda)$ obeys equations \eqref{eq:Vir_00_lambda}, \eqref{eq:Vir_11_lambda} and \eqref{eq:Vir_01_lambda}. Thus, it is related to $\hat{V}_j^3(\lambda)$ with a global rotation. The corresponding rotation matrix $M$ obeys
\begin{align}
M\left(\lambda\right)M\left(1/\lambda\right)&=I,\\
M^T\left(\lambda\right)M\left(\lambda\right)&=I,\\
\bar{M}\left(\bar{\lambda}\right)&=M\left(\lambda\right).
\end{align}
In any case, given a specific seed solution, one will be able to specify the matrix M\footnote{In the case of the BMN particle and the elliptic strings $M=-J$.}.
Similarly, equation \eqref{eq:t_reality} and the fact that the seed solution is a real function of $m_+$ and $m_-$ implies that
\begin{equation}\label{eq:V_Reality}
\bar{\hat{V}}\left(\bar{\lambda}\right)=\hat{V}\left(\lambda\right).
\end{equation}
It is trivial to show that the above imply
\begin{align}
m_1(\lambda)&=\left[C^T(\lambda)C(\lambda)\right]^{-1},\label{eq:def_m1_C}\\
m_2(\lambda)&=-C^{-1}(\lambda)M(\lambda)C(1/\lambda)J,\label{eq:def_m2_C}\\
m_3(\lambda)&=C^{-1}(\lambda)\bar{C}(\bar{\lambda}),\label{eq:def_m3_C}
\end{align}
These matrices satisfy identically \eqref{eq:constraint_m_inv}, \eqref{eq:constraint_m_cos} and \eqref{eq:constraint_m_real}. Furthermore, as for $\lambda=0$ the matrix $\hat{V}(\lambda)$ satisfies
\begin{equation}
\hat{V}(0)=U^T,
\end{equation}
it is evident that
\begin{equation}
C(0)=I.
\end{equation}
The latter implies that the equations \eqref{eq:constraint_m_0} are satisfied too.

The aftermath of this analysis is an unexpected statement. If one knows not only the seed solution, but also the whole family of solutions that correspond to the same Pohlmeyer counterpart as the seed\footnote{This is the case when one constructs NLSM via the ``inversion'' of the Pohlmeyer reduction in the spirit of \cite{Bakas:2016jxp}.}, then one can construct \emph{algebraically} the corresponding solution of the auxiliary system. For real values of the spectral parameter, the elements of the auxiliary field are constructed via an interpolation between different members of this family of solutions. In general they are determined by the analytic continuation of the family. 

In appendix \ref{subsec:simplest_dressing_factor} the simplest dressing factor is constructed. This contains a pair of poles on the unit circle at $e^{\pm i\theta_1}$. In appendix \ref{subsec:dressed_solution} we derive the corresponding dressed solution for a general seed. We show that this obeys the equations of motion and the same Virasoro constraints as the seed. The cosine of the Pohlmeyer field of the dressed solution is given by 
\begin{equation}\label{eq:add_cos_pohl}
m_+m_-\cos a^\prime=m_+m_-\cos a+\partial_+\partial_-\ln\left[\left(\hat{W}^T X_0\right)^2\right],
\end{equation}
where $\hat{W}^T=J \hat{V}(e^{i\theta_1})p$ and $p$ is a constant complex column, which obeys appropriate conditions (see appendix \ref{subsec:simplest_dressing_factor}). The argument of the logarithm is simply a linear combination of $\hat{V}^3_j$, i.e. the analytic continuation of the the family of the seed string solution. 

Notice that the proofs of the fact that the dressing trasformation with the simplest dressing factor preserves the Virasoro constraints and that the dressed solution obeys the equation of motion, are valid for any number of dimensions. Similarly, the structure of the addition formula \eqref{eq:add_cos_pohl} is the same for any NLSM defined on $\textrm{R}\times \textrm{S}^d$. Obviously, if  $d\geq3$, one has to appropriately generalize the presented solution of the auxiliary system $\hat{V}(\lambda)$.
\section{Discussion}
\label{sec:discussion}
Integrability of NLSMs on symmetric spaces stems from the existence of the Lax connection, which is flat and leads to an infinite tower of conserved charges. Yet, there are more aspects of integrability related to NLSMs. Given a seed solution of the NLSM, the dressing method enables the construction of new solutions of the NLSM through a pair of first order differential equations, the auxiliary system. Once this system is solved, multiple dressing transformations can be performed systematically. The Pohlmeyer reduction reveals that the embedding of the world-sheet into the target space, which is in turn embedded into a flat enhanced space is described by integrable models. Given a solution of the Pohlmeyer reduced theory, \Backlund transformations can be employed  in order to construct new solutions. Moreover, by substituting a solution of the Pohlmeyer reduced theory in the equations of motion of the NLSM, these become linear, since the Lagrange multiplier acts as a self-consistent potential. The dressing method and the \Backlund transformations are interrelated, as the application of the dressing method on the NLSM automatically performs a \Backlund transformation on the Pohlmeyer reduced theory. 

In this work we discussed strings, which, as time flows, propagate on a two-dimentional sphere. Their motion is described by the NLSM on $\mathrm{S}^2$. It is well known that the Pohlmeyer reduced theory of this NLSM is the sine-Gordon equation. We applied the dressing method on this NLSM using a mapping of $\mathrm{S}^2$ to the coset $\mathrm{SO}(3)/\mathrm{SO}(2)$. {Taking advantage of the parametrization introduced in \cite{Katsinis:2018ewd},\emph{ we obtained the solution of the auxiliary system for an arbitrary seed solution. This solution is built by combining appropriately the seed solution with a virtual one.} The latter has the same Pohlmeyer counterpart as the seed solution, it solves the NLSM equations of motion, yet, in general it is complex and obeys altered Virasoro constraints, which do not correspond to a valid string solution in $\mathbb{R}\times \mathrm{S}^2$. \emph{This virtual solution can be constructed trivially as long as one knows the whole class of solutions of the NLSM that correspond to a given solution of the Pohlmeyer reduced theory.} Subsequently, we constructed the solution of the NLSM that corresponds to the simplest dressing factor, namely the one that has a pair of poles on the unit circle. \emph{The dressed solution of the NLSM is a non-linear superposition of the seed solution of the NLSM and the virtual one.} This is a completely novel aspect of integrability of NLSMs.

Furthermore, we derived an addition formula for the on-shell Lagrangian density. This addition formula encapsulates the pair of the first order equations that constitute the \Backlund transformation of the sine-Gordon equation. We specify the relation between the location of the poles of the dressing factor and the spectral parameter of the \Backlund transformations. Our construction proves that the knowledge of the whole class of solutions of the NLSM that correspond to a given solution of the sine-Gordon equation, enables the insertion of solitons in this solution of the sine-Gordon equation \emph{without solving the equations of the \Backlund transformation.} As we obtained the general solution of the auxiliary system, our work implies that \emph{the dressing method is actually implementing the non-linear superposition we presented.} At the level of the sine-Gordon equation, since \emph{solitons} are inserted through \Backlund transformations, we showed that they \emph{are the Pohlmeyer counterpart of the non-linear superposition at the level of NLSM.} It is worth noticing that this non-linear superposition does not rely on finite gap integration and explicit construction of solutions of the NLSM; it is a fundamental property.

Non-linear equations are characterized by the fact that one can not construct new solutions of them by forming linear combinations of known solutions. Yet, the fact that the solutions of the Pohlmeyer reduced theory render the equations of motion of the NLSM linear seems to be a key element in our construction. Two solutions of the same equations of motion, which are effectively linear for the given solution of the sine-Gordon equation, are the ones that are superimposed in order to obtain a new solution. This operation constructs a dressed solution that does not correspond to the same Pohlmeyer field, thus the dressed solution belongs to a different ``effectively linearised" sector. It is interesting that starting from an arbitrary seed, the whole tower of sectors that are reached through the non-linear superposition is built by inserting solitons in the Pohlmeyer counterpart of the seed solution.

On the converse root, let us consider our construction from the point of view of the sine-Gordon equation. In order to perform a \Backlund tranformation on a given seed solution one needs to solve a pair of first order \emph{non-linear} differential equations. The presented analysis shows that this is equivalent to the construction of family of the NLSM solution, which correspond to the specific Pohlmeyer field. This requires the general solution of a \emph{linear} second order differential equation \eqref{eq:NLSM_EOM}, whereas the non-linear part of the calculation has become purely algebraic. The latter is the enforcement of the geometric and Virasoro constraints. Once the family has been constructed, the application of the dressing transformation using our construction and equation \eqref{eq:add_cos_pohl} effectively linearizes the \Backlund transformations.

The generalization of this work in symmetric spaces such as $\textrm{S}^d$, $\textrm{AdS}_d$, $\textrm{dS}_d$ and $\textrm{CP}^d$, as well as direct products of them, which are relevant for the gauge gravity duality, is highly interesting. The same is true for euclidean NLSMs on $\textrm{H}^d$ that are relevant for the holographic calculation of  Wilson Loops. In a similar manner one may study the implications of the choice of the coset that is used in order to implement the dressing method. It is known that the mapping of a symmetric space to different cosets may lead to different solutions \cite{Kalousios:2006xy}.

Finally, the implications of this construction to the physics of the NLSMs deserves a thorough study. Our previous works \cite{Katsinis:2019oox,Katsinis:2019sdo} revealed that dressed elliptic strings have interesting physical properties. A compelling finding is that there is a special class of dressed string solutions, which consists of the strings that correspond to the unstable modes of their precursors. These instabilities are related to the propagation of superluminal solitons on the background of the Pohlmeyer counterpart of the seed. It would be interesting to investigate whether one can discuss similar properties for arbitrary seed solutions in the context of the presented construction. Another potential implication of this construction regards the spectral problem of AdS/CFT. The latter was solved in \cite{Beisert:2005bm} in the thermodynamic limit, which is associated with long strings. Long strings can naturally be constructed via the application of the dressing method. Since they propagate on an infinite size world-sheet, the Pohlmeyer counterpart has a diverging period. The latter corresponds precisely to the existence of a soliton. As a result, long strings can be described as the non-linear superposition of a short string with a virtual one. It is interesting to study applications of our construction in this context, as it can be used for a general short string seed. As a last comment, the presented construction, in particular the addition formula for the cosine of the Pohlmeyer field, describes the instanton contributions to the action of the $\textrm{O}(3)$ sigma model over any zero instanton classical configuration. Maybe this could be incorporated for investigations along the lines of \cite{Krichever:2020tgp}.

\subsection*{Acknowledgements}
The research of D.K. is co-financed by Greece and the European Union (European Social Fund- ESF) through the Operational Programme ``Human Resources Development, Education and Lifelong Learning'' in the context of the project ``Strengthening Human Resources Research Potential via Doctorate Research'' (MIS-5000432), implemented by the State Scholarships Foundation (IKY). The research of G.P. has received funding from the Hellenic Foundation for Research and Innovation (HFRI) and the General Secretariat for Research and Technology (GSRT), in the framework of the ``First Post-doctoral researchers support'', under grant agreement No 2595.

\appendix
\section{Implications of Orthonormality}\label{sec:Normalization}
In this section we enforce the condition that $\vec{\hat{V}}_i$ should form an orthonormal basis as suggested by \eqref{eq:con_orthonormal}. We first discuss the implication of the normalization. Using \eqref{eq:column_form}, the relevant constraints are
\begin{equation}\label{eq:eq_norm}
\frac{\vert\vec{\tau}_1\vert^2\left(\partial_0 \hat{V}_j^3\right)^2 +\vert\vec{\tau}_0\vert^2\left(\partial_1 \hat{V}_j^3\right)^2-2\vec{\tau}_0\cdot\vec{\tau}_1\left(\partial_0 \hat{V}_j^3\right)\left(\partial_1 \hat{V}_j^3\right)}{\vert\vec{\tau}_0\times\vec{\tau}_1\vert^2}+\left(\hat{V}_j^3\right)^2=1.
\end{equation}
For $j=3$ one obtains the following equation
\begin{equation}
\vert\vec{\tau}_1\vert^2\left(\partial_0 \Theta\right)^2 +\vert\vec{\tau}_0\vert^2\left(\partial_1  \Theta\right)^2-2\vec{\tau}_0\cdot\vec{\tau}_1\left(\partial_0  \Theta\right)\left(\partial_1  \Theta\right)=\vert\vec{\tau}_0\times\vec{\tau}_1\vert^2.\label{eq:eq_norm_theta}
\end{equation}
Using this result, the $j=1$ and $j=2$ equations imply
\begin{align}
&\sin^2\Theta\left[\vert\vec{\tau}_1\vert^2\left(\partial_0 \Phi\right)^2 +\vert\vec{\tau}_0\vert^2\left(\partial_1  \Phi\right)^2-2\vec{\tau}_0\cdot\vec{\tau}_1\left(\partial_0  \Phi\right)\left(\partial_1  \Phi\right)\right]=\vert\vec{\tau}_0\times\vec{\tau}_1\vert^2,\label{eq:eq_norm_phi}\\
&\vert\vec{\tau}_1\vert^2\left(\partial_0 \Theta\right)\left(\partial_0 \Phi\right) +\vert\vec{\tau}_0\vert^2\left(\partial_1 \Theta\right)\left(\partial_1 \Phi\right)-\vec{\tau}_0\cdot\vec{\tau}_1\left[\left(\partial_0  \Theta\right)\left(\partial_1  \Phi\right)+\left(\partial_1  \Theta\right)\left(\partial_0  \Phi\right)\right]=0.\label{eq:eq_norm_hom1}
\end{align}
Equations \eqref{eq:eq_norm_theta} and \eqref{eq:eq_norm_phi} are equivalent to
\begin{align}
\begin{split}\label{eq:eq_norm_den}
&\vert\vec{\tau}_1\vert^2\left[\left(\partial_0\Theta\right)^2+\sin^2\Theta\left(\partial_0\Phi\right)^2\right] +\vert\vec{\tau}_0\vert^2\left[\left(\partial_1\Theta\right)^2+\sin^2\Theta\left(\partial_1\Phi\right)^2\right]\\&\qquad\qquad\qquad\qquad\qquad\qquad-2\vec{\tau}_0\cdot\vec{\tau}_1\left[\partial_0\Theta\partial_1\Theta+\sin^2\Theta\partial_0\Phi\partial_1\Phi\right]=2\vert \vec{\tau}_0\times\vec{\tau}_1\vert^2,
\end{split}\\
\begin{split}\label{eq:eq_norm_hom2}
&\vert\vec{\tau}_1\vert^2\left[\left(\partial_0\Theta\right)^2-\sin^2\Theta\left(\partial_0\Phi\right)^2\right] +\vert\vec{\tau}_0\vert^2\left[\left(\partial_1\Theta\right)^2-\sin^2\Theta\left(\partial_1\Phi\right)^2\right]\\&\qquad\qquad\qquad\qquad\qquad\qquad-2\vec{\tau}_0\cdot\vec{\tau}_1\left[\partial_0\Theta\partial_1\Theta-\sin^2\Theta\partial_0\Phi\partial_1\Phi\right]=0.
\end{split}
\end{align}
Equations \eqref{eq:eq_norm_hom1} and \eqref{eq:eq_norm_hom2} yield
\begin{align}
\vert\vec{\tau}_0\vert^2&=\frac{\left(\partial_0\Theta\right)^2+\sin^2\Theta\left(\partial_0\Phi\right)^2}{\partial_0\Theta\partial_1\Theta+\sin^2\Theta\partial_0\Phi\partial_1\Phi}\vec{\tau}_0\cdot\vec{\tau}_1,\\
\vert\vec{\tau}_1\vert^2&=\frac{\left(\partial_1\Theta\right)^2+\sin^2\Theta\left(\partial_1\Phi\right)^2}{\partial_0\Theta\partial_1\Theta+\sin^2\Theta\partial_0\Phi\partial_1\Phi}\vec{\tau}_0\cdot\vec{\tau}_1.
\end{align}
Finally, using the identity $\vert\vec{\tau}_0\times\vec{\tau}_1\vert^2= \vert\vec{\tau}_0\vert^2\vert\vec{\tau}_1\vert^2-\left(\vec{\tau}_0\cdot\vec{\tau}_1\right)^2$ and substituting these results into equation \eqref{eq:eq_norm} we specify $\vec{\tau}_0\cdot\vec{\tau}_1$ as  
\begin{equation}\label{eq:t01_equation}
\vec{\tau}_0\cdot\vec{\tau}_1=\partial_0\Theta\partial_1\Theta+\sin^2\Theta\partial_0\Phi\partial_1\Phi.
\end{equation}
As a consequence
\begin{equation}\label{eq:tisq_equation}
\vert\vec{\tau}_i\vert^2=\left(\partial_i\Theta\right)^2+\sin^2\Theta\left(\partial_i\Phi\right)^2.
\end{equation}
Having obtained these results, it is straightforward to show that the vectors $\vec{\hat{V}}_j$ and $\vec{\hat{V}}_k$ are orthogonal to each other for $j \neq k$.
\section{The Remaining Equation of the Auxiliary System}\label{sec:remaining}
In this appendix we show that the equations \eqref{eq:Vir_00_lambda}, \eqref{eq:Vir_11_lambda} and \eqref{eq:Vir_01_lambda} imply that the equation \eqref{eq:last_eq} for $j=1,2,3$ is satisfied without any further constraints on $\Theta$ and $\Phi$. This equation contains $\hat{V}_j^3,$ as well as derivatives of it. Since we do not rely on an explicit expression for the seed solution, we are not able to proceed directly, thus we will work with appropriate projections of this equation. The form of \eqref{eq:column_form} implies that
\begin{align}
\sum_j\left[\vec{\hat{V}}_j\times\vec{X}_0\right]\hat{V}_j^3&=0,\label{eq:sum_V}\\
\sum_j\left[\vec{\hat{V}}_j\times\vec{X}_0\right]\partial_0 \hat{V}_j^3&=\frac{\vert\vec{\tau}_0\vert^2}{\left(\vec{\tau}_0\times\vec{\tau}_1\right)\cdot\vec{X}_0}\vec{\tau}_1\times\vec{X}_0-\frac{\vec{\tau}_0\cdot\vec{\tau}_1}{\left(\vec{\tau}_0\times\vec{\tau}_1\right)\cdot\vec{X}_0}\vec{\tau}_0\times\vec{X}_0=\vec{\tau}_0,\label{eq:sum_V_d0}\\
\sum_j\left[\vec{\hat{V}}_j\times\vec{X}_0\right]\partial_1 \hat{V}_j^3&=\frac{\vec{\tau}_0\cdot\vec{\tau}_1}{\left(\vec{\tau}_0\times\vec{\tau}_1\right)\cdot\vec{X}_0}\vec{\tau}_1\times\vec{X}_0-\frac{\vert\vec{\tau}_1\vert^2}{\left(\vec{\tau}_0\times\vec{\tau}_1\right)\cdot\vec{X}_0}\vec{\tau}_0\times\vec{X}_0=\vec{\tau}_1,\label{eq:sum_V_d1}
\end{align}
where we used the equation \eqref{eq:con_orthonormal}, as well as the equations \eqref{eq:Vir_00_lambda}, \eqref{eq:Vir_11_lambda} and \eqref{eq:Vir_01_lambda}\footnote{Since $\vec{\tau}_i\cdot\vec{X}_0=0$ one can decompose $\vec{\tau}_i$ on the basis which consists of $\vec{\tau}_0\times\vec{X}_0$ and $\vec{\tau}_1\times\vec{X}_0$.}. Similarly one can obtain
\begin{align}
\sum_j \hat{V}_j^3\partial_i^2 \hat{V}_j^3&=-\vert\vec{\tau}_i\vert^2,\label{eq:sum_ddV}\\
\sum_j\partial_0 \hat{V}_j^3\partial_0^2 \hat{V}_j^3&=\vec{\tau}_0\cdot\partial_0\vec{\tau}_0,\quad \sum_j\partial_1 \hat{V}_j^3\partial_1^2 \hat{V}_j^3=\vec{\tau}_1\cdot\partial_1\vec{\tau}_1.\label{eq:sum_ddV_d}
\end{align}
The equations \eqref{eq:sum_V} and \eqref{eq:sum_ddV} suggest that the projection of \eqref{eq:last_eq} on $\hat{V}_j^3$ is satisfied identically. Likewise, equations \eqref{eq:sum_V_d0},  \eqref{eq:sum_V_d1} and \eqref{eq:sum_ddV_d} imply that the projections of \eqref{eq:last_eq} for $i=0$ on $\partial_0 \hat{V}_j^3$ and for $i=1$ on $\partial_1 \hat{V}_j^3$ are
\begin{align}
\vec{\tau}_0\cdot\partial_0\vec{\tau}_0&=\left(\partial_0\vec{\tau}_0-\vec{t}_0\times\vec{\tau}_0\right)\cdot\vec{\tau}_0,\\
\vec{\tau}_1\cdot\partial_1\vec{\tau}_1&=\left(\partial_1\vec{\tau}_1-\vec{t}_1\times\vec{\tau}_1\right)\cdot\vec{\tau}_1,
\end{align}
respectively. These equations are identically true.
\section{The Dressing Factor with a Pair of Poles}
\label{sec:dressing_factor}
\subsection{The Construction of the Dressing Factor}
\label{subsec:simplest_dressing_factor}
In this section we construct the simplest dressing factor. We begin the analysis by presenting the constraints that have to be imposed on the dressing factor
\begin{align}
\chi^{-1}(\lambda)&=\chi^T(\lambda),\label{eq:chi_inverse}\\
\chi\left(1/\lambda\right)&=g^\prime J\chi\left(\lambda\right)gJ,\label{eq:chi_coset}\\
\bar{\chi}\left(\bar{\lambda}\right)&=\chi\left(\lambda\right)\label{eq:chi_reality},
\end{align}
so that $\Psi^\prime(\lambda)=\chi(\lambda)\Psi(\lambda)$ obeys \eqref{eq:Psi_inv}, \eqref{eq:Psi_cos} and \eqref{eq:Psi_real}.

Considering meromorphic dressing factors, the above constraints naively suggest that the poles must form quadruplets of the form $(\lambda_1, \bar{\lambda}_1, \lambda_1^{-1}, \bar{\lambda}_1^{-1})$\cite{Miramontes:2008wt}. Thus, the dressing factor should have the following structure:
\begin{equation}
\chi\left(\lambda\right)=I+\frac{Q}{\lambda-\lambda_1}+\frac{\bar{Q}}{\lambda-\bar{\lambda_1}}+\frac{\tilde{Q}}{\lambda-\lambda_1^{-1}}+\frac{\bar{\tilde{Q}}}{\lambda-\bar{\lambda_1}^{-1}}.
\end{equation}
Concerning the discussion about the reduction group, the reader should notice that for $\lambda=0$, the equations \eqref{eq:chi_inverse}, \eqref{eq:chi_coset} and \eqref{eq:chi_reality} reduce to 
\begin{align}
\bar{\chi}\left(0\right)&=\chi\left(0\right),\\
\chi^{-1}(0)&=\chi^T(0),\\
I&=g^\prime \theta g^\prime\theta.
\end{align}
As long as $\Psi(0)=g$ these relations imply that $g^\prime$ satisfies \eqref{eq:g_constraints_1} and \eqref{eq:g_constraints_2} and thus, it belongs to the coset $\textrm{SO}(3)/\textrm{SO}(2)$. In the following, we show that the precise form of the matrices $m_i(\lambda)$ is irrelevant.

We restrict our analysis to the simplest dressing factor. This emerges, when the poles lie on the unit circle, i.e.
\begin{equation}
\lambda_1= e^{i \theta_1},
\end{equation}
where $\theta_1\in\mathbb{R}.$ This choice implies that the location of the poles at $\bar{\lambda}_1$ and $\lambda_1^{-1}$ coincide and thus, the quadruplet of poles reduces to a doublet. After an appropriate redefinition of the residues, the dressing factor can be expressed as
\begin{equation}\label{eq:chi_2_poles}
\chi=I + \frac{e^{i \theta_1}-e^{-i \theta_1}}{\lambda-e^{i \theta_1}}Q-\frac{e^{i \theta_1}-e^{-i \theta_1}}{\lambda-e^{-i \theta_1}}\bar{Q}.
\end{equation}
Clearly, the constraint \eqref{eq:chi_reality} is satisfied. We postulate that the inverse of the dressing factor is given by \eqref{eq:chi_inverse}. Next, we impose the relation $\chi\chi^{-1}=I.$ The cancellation of the residue of the second order poles at $e^{i \theta_1}$ and $e^{-i \theta_1}$ requires $QQ^T=0$. Then, the cancellation of the residues of the first order poles on the same locations implies that
\begin{equation}
Q\left(I -Q^\dagger\right)+\left(I -\bar{Q}\right)Q^T=0.
\end{equation}
Clearly, this relation is satisfied if
\begin{equation}
Q=Q^\dagger
\end{equation}
and $Q$ is a projection matrix, i.e. it obeys $Q^2=Q$. The relation $\chi^{-1}\chi=I$ implies $Q^TQ=0$.
We define
\begin{equation}
Q=\frac{F F^\dagger}{F^\dagger F}.
\end{equation}
where $F$ is a vector. The constraints $QQ^T=Q^TQ=0$ imply
\begin{equation}
F^TF=0.\label{eq:constraint_inv_2nd_order}
\end{equation}
The requirement that $\Psi^\prime$ satisfies the auxiliary system implies the equations of motion of the dressing factor, which read
\begin{equation}\label{eq:chi_eom}
\left(1\pm\lambda\right)\left(\partial_\pm \chi\right)\chi^{-1}+\chi\left(\partial_\pm g\right)g^{-1}\chi^{-1}=\left(\partial_\pm g^\prime\right)g^{\prime-1}.
\end{equation}
For $\lambda=0$ these equations are satisfied trivially, thus one needs only to ensure that the residues of the various poles cancel. The right-hand-side of \eqref{eq:chi_eom} does not depend on $\lambda,$ thus, the same must hold true for the left-hand-side. The cancellation of the residues of the second order poles at $e^{i \theta_1}$ suggests that
\begin{equation}\label{eq:F_eom}
\left(1\pm e^{i \theta_1}\right)\partial_\pm F^\dagger+F^\dagger\left(\partial_\pm g\right)g^{-1}=0.
\end{equation}
This equation implies that
\begin{equation}
F^\dagger=p^\dagger \Psi^{-1}(e^{i \theta_1}),\label{eq:eom_F}
\end{equation}
where $p$ is a constant vector. Taking into account \eqref{eq:def_m1_C} and \eqref{eq:def_m3_C} it is easy to show that
\begin{equation}
F=\Psi(e^{-i\theta_1})\left[C^\dagger(e^{i\theta_1})C(e^{-i\theta_1})\right]^{-1}p.\label{eq:def_F}
\end{equation}
It is a matter of elementary algebra to show that
\begin{equation}
F^TF=p^T\left[C^\dagger(e^{i\theta_1})\bar{C}(e^{i\theta_1})\right]^{-1}p.
\end{equation}
We may redefine the constant vector $p$ as follows
\begin{equation}
p=C^\dagger(e^{i\theta_1})\tilde{p},
\end{equation}
where 
\begin{equation}
\tilde{p}^T\tilde{p}=0\label{eq:projector_null},
\end{equation}
in order to satisfy \eqref{eq:constraint_inv_2nd_order}. The equation \eqref{eq:def_F} may be re-expressed as 
\begin{equation}
F=\Psi(e^{-i\theta_1})C^{-1}(e^{-i\theta_1})\tilde{p}=V(e^{-i\theta_1})\tilde{p},
\end{equation}
where
\begin{equation}
V(\lambda)=J U J \hat{V}(\lambda).
\end{equation}
Furthermore, in view of \eqref{eq:def_m2_C} we may obtain
\begin{equation}
gJ F=-\Psi(e^{i\theta_1})C^{-1}(e^{i\theta_1})M(e^{i\theta_1})\tilde{p}
\end{equation}
and 
\begin{equation}
\bar{F}=\Psi(e^{i\theta_1})C^{-1}(e^{i\theta_1})\bar{\tilde{p}}.
\end{equation}
Clearly, if
\begin{equation}\label{eq:projector_complex_conj}
\bar{\tilde{p}}=-M(e^{i\theta_1})\tilde{p},
\end{equation}
we will obtain
\begin{equation}
\bar{F}=gJ F,\label{eq:F_bar_coset}
\end{equation}
which implies
\begin{equation}
\bar{Q}=gJ Q gJ.
\end{equation}
The above formulae and equation \eqref{eq:definition_C} imply that the precise form of the matrix $C$ does not affect $F$, in the sense that $C$ can be absorbed into the selection of the constant vector $p$. Without loss of generality, it may be substituted by $C(0)=I$. Moreover, for $M=-J$, the equations \eqref{eq:projector_null} and \eqref{eq:projector_complex_conj} restrict the constant vector $\tilde{p}$ to be of the form
\begin{equation}\label{eq:projector_form}
\tilde{p}=\begin{pmatrix}
\cos w \\
\sin w \\
i
\end{pmatrix},
\end{equation}
where $w\in\mathbb{R}$. The overall scale is irrelevant since it cancels at the the level of the residues $Q$.

Then, equating the residues of the left-hand-side and right-hand-side of \eqref{eq:chi_coset} we obtain
\begin{equation}
\bar{Q}=e^{-2 i \theta_1}g^\prime J Qg\theta ,
\end{equation}
while the analytic part implies
\begin{equation}\label{eq:chi_coset_analytical}
\chi(0)g J\chi(0)=g J.
\end{equation}
This relation is the coset constraint $g^\prime J g^\prime J=I$. It is simple to show that these relations are indeed satisfied. Finally, it is a matter of algebra to show that the residues of the first order poles of the equations of motion \eqref{eq:chi_eom} cancel.

Even though the precise form of $M$ is not important for the results of this paper, in the general case it is of the form 
\begin{equation}\label{eq:M_decomposition}
M(\lambda)=U^\dagger(\lambda)\left(-J\right)U(\lambda),
\end{equation}
where
\begin{align}
U\left(\lambda\right)U\left(1/\lambda\right)&=I,\\
U^T\left(\lambda\right)U\left(\lambda\right)&=I,\\
\bar{U}\left(\bar{\lambda}\right)&=U\left(\lambda\right).
\end{align}
Equation \eqref{eq:m_of_lambda} implies that the inversion $\lambda\rightarrow1/\lambda$ is equivalent to the transformation $m_\pm\rightarrow-m_\pm$. This transformation leaves the equations of motion, as well as the Virasoro constraints and the geometric constraint, invariant. Thus the inversion of $\lambda$ can at most perform an $\mathrm{SO} \left( 3 \right)$ transformation on the solution. The form of the Virasoro constraints suggests that the transformation $m_\pm\rightarrow-m_\pm$ is equivalent to the reflection of the worldsheet coordinates, i.e. it can be perceived as an inversion of the vectors $\partial_\pm X$ in the enhanced space. In the ``hatted'' frame, where $X$ is expressed as a rotation acting on $X_0$, the directions transverse to the latter, should be inverted, up to a global rotation, i.e. the matrix $M$ cannot be e.g. the identity matrix, but rather it is of the form \eqref{eq:M_decomposition}. In this general case, equation \eqref{eq:M_decomposition} implies that it is the vector $U(e^{i\theta_1})\tilde{p}$ which is of the form \eqref{eq:projector_form}.

\subsection{The Dressed Solution for a Pair Poles on the  Unit Circle}
\label{subsec:dressed_solution}
Using the dressing factor, which is constructed in appendix \ref{subsec:simplest_dressing_factor}, the dressed element of the coset reads
\begin{equation}\label{eq:gprime_psi}
g^\prime=J\left[J g - \frac{e^{i \theta_1}-e^{-i \theta_1}}{e^{i \theta_1}}J g\frac{W W^T}{W^T J g W}J g+\frac{e^{i \theta_1}-e^{-i \theta_1}}{e^{-i \theta_1}}\frac{W W^T}{W^T J g W}\right],
\end{equation}
where $W$ is given by
\begin{equation}\label{eq:def_W}
W=J\bar{F}.
\end{equation}
Notice that $W^T W=0$ as a direct consequence of \eqref{eq:constraint_inv_2nd_order}. Using the mapping \eqref{eq:g_mapping}, along with the last relation, we obtain 
\begin{equation}
X^\prime=\cos\theta_1 X+i \sin\theta_1\left(\frac{W}{W^TX}-X\right),
\end{equation}
where $X^\prime$ corresponds to the element of the coset $g^\prime$. Notice that $X^TX^\prime=\cos\theta_1$, which implies that each point of the dressed string lies on an epicycle of angle $\theta_1$, which is centered at a point of the seed string\cite{Katsinis:2018ewd}. In order to express the above relation in a manifestly real form we implement \eqref{eq:F_bar_coset}. The latter in view of \eqref{eq:g_mapping} implies
\begin{equation}
X=\frac{W -\bar{W}}{2W^TX}\Rightarrow X^T\bar{W}=-X^TW.
\end{equation}
Thus, the dressed solution of the NLSM reads
\begin{equation}
X^\prime=\cos\theta_1 X+ \sin\theta_1 X_w,
\end{equation}
where
\begin{equation}
X_w=i \frac{W +\bar{W}}{X^T\left(W -\bar{W}\right)}=\frac{\re{W}}{X^T\im{W}}
\end{equation}
is a unit norm vector, which is perpendicular to $X$.

Let us use the specific form of $X^\prime$ in order to verify that it satisfies same the Virasoro constraints as the seed, the NLSM equations of motion and specify the corresponding Pohlmeyer field and its connection to that of the seed. Taking into account \eqref{eq:def_W}, the equations of motion \eqref{eq:F_eom} imply
\begin{equation}
\partial_\pm W=\frac{1}{1\pm e^{i\theta_1}}J\left(\partial_\pm g\right)J g W.
\end{equation} 
Substituting $g$ and using the mapping \eqref{eq:g_mapping}, we obtain
\begin{equation}\label{eq:W_der}
\partial_\pm W=\frac{2}{1\pm e^{i\theta_1}}\left(\left(X^TW\right)\partial_\pm X-\left( W^T\partial_\pm X\right)X\right).
\end{equation}
In addition, since $X$ is unit norm, it follows that
\begin{equation}\label{eq:WX_der}
\partial_\pm\left( W^TX\right)=-\frac{1\mp e^{i\theta_1}}{1\pm e^{i\theta_1}}W^T\partial_\pm X.
\end{equation}
Substituting the latter into \eqref{eq:W_der} yields
\begin{equation}
\partial_\pm W=\frac{2}{1\pm e^{i\theta_1}}\left(X^TW\right)\partial_\pm X+\frac{2}{1\mp e^{i\theta_1}}\partial_\pm\left( W^TX\right)X.
\end{equation}
Taking the complex conjugate and using the fact that $X^T\bar{W}=-X^TW$, we obtain 
\begin{equation}
\partial_\pm \bar{W}=-\frac{2}{1\pm e^{-i\theta_1}}\left(X^TW\right)\partial_\pm X-\frac{2}{1\mp e^{-i\theta_1}}\partial_\pm\left( W^TX\right)X.
\end{equation} 
The above imply that
\begin{equation}
\partial_\pm \left(W+\bar{W}\right)=2\frac{1\mp e^{i\theta_1}}{1\pm e^{i\theta_1}}\left(X^TW\right)\partial_\pm X+2\frac{1\pm e^{i\theta_1}}{1\mp e^{i\theta_1}}\left(\partial_\pm\left(W^TX\right)\right)X.
\end{equation}
Taking everything into account, we obtain
\begin{equation}
\partial_\pm X^\prime=\pm\left(\partial_\pm X-\frac{\partial_\pm\left(W^TX\right)}{W^TX}X\right)-\frac{\partial_\pm\left(W^TX\right)}{W^TX}X^\prime.
\end{equation}
Having this equation at hand it is possible to proceed into the necessary verifications.

It is a matter of algebra to show that
\begin{equation}
\left(\partial_\pm X^\prime\right)^T\left(\partial_\pm X^\prime\right)=\left(\partial_\pm X\right)^T\left(\partial_\pm X\right)=m^2_\pm,
\end{equation}
thus the Virasoro constraints are preserved by the dressing transformation. It order to derive the above it is advantageous to use the relation $\left(\partial_\pm X\right)^TX^\prime=-X^T\left(\partial_\pm X^\prime\right)$. Similarly, one can show that the cosine of the dressed Pohlmeyer field is related to the one of the seed as
\begin{equation}\label{eq:pohl_addition}
\left(\partial_+ X^\prime\right)^T \partial_- X^\prime=-\left(\partial_+ X\right)^T\partial_- X-2\frac{\partial_+\left(W^TX\right)\partial_-\left(W^T X\right)}{\left(W^T X\right)^2}.
\end{equation}
Using \eqref{eq:W_der} and \eqref{eq:WX_der} it is easy to show that
\begin{equation}\label{eq:WX_der2}
\partial_+\partial_-\left(W^T X\right)=-\left[\left(\partial_+ X\right)^T \partial_- X\right]\left(W^T X\right).
\end{equation}
It is a matter of algebra to show that
\begin{multline}
\partial_+\partial_-X^\prime+\left[\left(\partial_+ X^\prime\right)^T \partial_- X^\prime\right]X^\prime \\
=\partial_+\partial_-X^\prime+\left[-\left(\partial_+ X\right)^T \partial_- X-2\frac{\partial_+ \left(W^T X\right)\partial_-\left(W^T X\right)}{\left(W^T X\right)^2}\right]X^\prime =\\
\partial_+\partial_-X+\left[\left(\partial_+ X\right)^T \partial_- X\right]X,
\end{multline}
which proves that the equations of motion for $X^\prime$ are satisfied as long as the ones for $X$ do so.

The identity
\begin{equation}
\frac{\partial_+f\partial_-f}{f^2}=\frac{\partial_+\partial_-f}{f}-\partial_+\partial_-\ln f
\end{equation}
and \eqref{eq:WX_der2} imply that \eqref{eq:pohl_addition} assumes the form
\begin{equation}\label{eq:algebraic_addition_1}
\left(\partial_+ X^\prime\right)^T \partial_- X^\prime=\left(\partial_+ X\right)^T \partial_- X+\partial_+\partial_-\ln\left[\left(W^T X\right)^2\right].
\end{equation}
This is an algebraic addition formula for the cosine of the Pohlmeyer field. In the context of AdS/CFT, the latter is the on-shell Lagrangian density of the $\mathrm{S}^2$ part of the string action.

The sine of the Pohlmeyer field of the dressed solution is given by
\begin{multline}
m_+m_-\sin a^\prime=X^\prime\cdot\left(\partial_+ X^\prime\times\partial_- X^\prime\right)=-\cos\theta_1 X\cdot\left(\partial_+ X\times\partial_- X\right)\\-\sin\theta_1\left[\frac{\partial_+ \left(W^T X\right)}{\left(W^T X\right)} X\cdot\left(X_w\times\partial_- X\right)+\frac{\partial_- \left(W^T X\right)}{\left(W^T X\right)} X\cdot\left(\partial_+ X\times X_w\right)\right].
\end{multline}
One can easily show that
\begin{equation}
X_w^T\partial_\pm X=-\frac{X^T\partial_\pm X^\prime}{\sin\theta_1}=\pm\frac{\partial_\pm \left(W^T X\right)}{\left(W^T X\right)}\frac{1\pm \cos\theta_1}{\sin\theta_1},
\end{equation}
which allows the expansion of $X_w$ in the basis formed by the vectors $\partial_\pm X$. After some tedious, but straightforward, algebra one may obtain
\begin{multline}
\sin a\sin a^\prime=-\cos\theta_1 \left[1+\cos a\cos a^\prime\right]\\-\left[\left(\frac{\partial_+ \left(W^T X\right)}{m_+\left(W^T X\right)}\right)^2 \left(1+\cos\theta_1\right)-\left(\frac{\partial_- \left(W^T X\right)}{m_-\left(W^T X\right)}\right)^2 \left(1-\cos\theta_1\right)\right],
\end{multline}
which is equivalent to
\begin{multline}
\left[\cos\left(\frac{a-a^\prime}{2}\right)\right]^2\cot\left(\frac{\theta_1}{2}\right)-\left[\cos\left(\frac{a+a^\prime}{2}\right)\right]^2\tan\left(\frac{\theta_1}{2}\right)=\\
\left(\frac{\partial_- \left(W^T X\right)}{m_-\left(W^T X\right)}\right)^2 \tan\left(\frac{\theta_1}{2}\right)-\left(\frac{\partial_+ \left(W^T X\right)}{m_+\left(W^T X\right)}\right)^2 \cot\left(\frac{\theta_1}{2}\right).
\end{multline}
In addition, \eqref{eq:pohl_addition} can be written in the form
\begin{equation}
\cos\left(\frac{a-a^\prime}{2}\right)\cos\left(\frac{a+a^\prime}{2}\right)=-\left(\frac{\partial_+ \left(W^T X\right)}{m_+\left(W^T X\right)}\right)\left(\frac{\partial_- \left(W^T X\right)}{m_-\left(W^T X\right)}\right).
\end{equation}
Thus, it is trivial to show that
\begin{align}
\frac{\partial_+ \left(W^T X\right)}{\left(W^T X\right)}&=\pm m_+\tan\left(\frac{\theta_1}{2}\right)\cos\left(\frac{a+a^\prime}{2}\right),\\
\frac{\partial_- \left(W^T X\right)}{\left(W^T X\right)}&=\mp m_-\cot\left(\frac{\theta_1}{2}\right)\cos\left(\frac{a-a^\prime}{2}\right).
\end{align}
Finally, \eqref{eq:algebraic_addition_1} implies that
\begin{equation}
\partial_+\partial_-\ln\left(W^T X\right)=m_+m_-\sin\left(\frac{a+a^\prime}{2}\right)\sin\left(\frac{a-a^\prime}{2}\right).
\end{equation}
Since $\partial_+\partial_-\ln\left(W^T X\right)=\partial_-\partial_+\ln\left(W^T X\right)$, the latter corresponds to the pair of first order equations, which are the \Backlund transformations of the sine-Gordon equation \eqref{eq:SG}. These read
\begin{align}
\partial_+\left(\frac{a-a^\prime}{2}\right)&=\alpha  m_+\sin\left(\frac{a+a^\prime}{2}\right),\\
\partial_-\left(\frac{a+a^\prime}{2}\right)&=-\frac{1}{\alpha} m_-\sin\left(\frac{a-a^\prime}{2}\right),
\end{align}
where, according to the presented analysis, the parameter $\alpha$ of the \Backlund transformation is given by
\begin{equation}
\alpha=\pm\tan\left(\frac{\theta_1}{2}\right).
\end{equation}
The sign of $\alpha$ can not be determined since it corresponds to the freedom of shifting either $a$ or $a^\prime$ by $2\pi$.

\end{document}